\documentclass[aps,pre,twocolumn,showpacs]{revtex4-1}

\usepackage{amsmath,amssymb,amsfonts}

\usepackage{inputenc}
\usepackage{subfigure}
\usepackage{graphicx}

\newcommand{\average}[1]{ \left< #1 \right>}
\newcommand{\avg}[1]{ \langle #1 \rangle}
\newcommand{\avgt}[1]{\langle #1 \rangle_t}

\begin{document}
\title{Noise-induced volatility of collective dynamics}

\author{Georges Harras$^1$}
\author{Claudio J. Tessone$^2$}
\author{Didier Sornette$^{1}$}
\affiliation{$^1$ Chair of Entrepreneurial Risks, D-MTEC,
ETH Zurich, CH-8092 Zurich, Switzerland}
\affiliation{$^2$ Chair of Systems Design, D-MTEC,
ETH Zurich, CH-8092 Zurich, Switzerland}
\begin{abstract}
{\em Noise-induced volatility} 
refers to a phenomenon of  increased level of fluctuations in the collective dynamics of bistable units
in the presence of a rapidly varying external signal, and intermediate noise levels. The archetypical signature of this phenomenon
is that --beyond the increase in the level of fluctuations-- the response of the system becomes uncorrelated with the external driving force, making it different from stochastic resonance.
Numerical simulations and an analytical theory of a stochastic
dynamical version of the Ising model on regular and random networks
demonstrate the ubiquity and robustness of this phenomenon, which is argued to be
a possible cause of excess volatility in financial markets, of enhanced
effective temperatures in a variety of out-of-equilibrium systems and of
strong selective responses of immune systems of complex biological organisms.
 Extensive numerical simulations are compared with a mean-field theory for different
 network topologies.
\end{abstract}
\date{\today}
\maketitle

\section{Introduction}

Noise  has effects {\em a priori} unexpected  on the organization of complex systems made of interacting elements,
 as shown by  {\em stochastic resonance} (SR) \cite{GHJM:1998}, {\em coherence resonance} \cite{PK:1997}, {\em noise-induced
 phase transitions} \cite{BPT:1994}, noise-induced transport 
 \cite{HM:2009} and its game theoretical version, the {\em Parrondo's Paradox} \cite{Abbott2010Asymmetry}.
 SR occurs in a system when a small applied (sub-threshold) periodic signal is amplified by the
 addition of noise and the maximum of amplification is found for intermediate noise strengths.
 More generally, SR refers to the situation where
 noise and nonlinearity combine to increase the strength in the system response.
 Among others, SR was shown to appear in optical \cite{Mcnamara1988Observation}
 and magnetic systems \cite{Leung1998Response,Leung1999Nontrivial}, thought to be relevant from
 the Earth climate \cite{Nicolis1993SRclimateChange} and the dynamics of ice ages \cite{Benzi1983SRClimateChange},
 to neurobiology \cite{Douglass1993SRcrayfish,Levin1996SRcricket} and visual perceptions 
 \cite{Simonotto1997SRvisualperc}.
 Generally SR is studied in bistable systems,
 where the amplification of a subthreshold periodic signal is achieved through the synchronization of
 noise-induced inter-well hopping of the dynamic variable and the driving signal.
 The signal is maximally amplified when the level of noise is such that the Kramers time, which
 is the intrinsic lifetime associated with the noise-induced transition between the two stable states,
 equals half of the period of the external forcing. 
 The bistability of the dynamic variable
 can be given either explicitly \cite{Gammaitoni1995Stochastic} --for low dimensional systems-- 
 or emerge from the interaction of the many constituents as for the magnetization in the Ising model
 in the ferromagnetic case \cite{Acharyya2006Nonequilibrium,Korniss2002Absence}.
 However, systems of many interacting constituents may depart from the paradigmatic setting of SR,
 as there is no inter-well hopping for the macroscopic observable,  and thus the bistability is
 only preserved at the microscopic level.

The Ising model, driven by a periodic signal, has been extensively studied in the realm of the kinetic Ising
 model and dynamical phase transitions
\cite{TomeOliveira1990,Korniss2002Absence,Buendia2008Dynamic,Sides1998Kinetic,Acharyya2006Nonequilibrium}.
 When a spatially extended Ising system --for temperatures below the Curie temperature-- is
 forced by a weak periodic influence, the
 magnetization performs dynamics around its nonzero equilibrium fixed point, resulting
 in a nonzero time-averaged magnetization.
 For a given temperature, by increasing either the field strength or the period of the signal, the system becomes able to hop between
 the two symmetric equilibrium fixed points inducing a zero average magnetization.
 The nature of this transition, from a nonzero to a zero average magnetization,
 can be manifold and depends on the control parameter.
 For weak, subthreshold forcing strength and finite-size systems --where inter-well hopping is enabled
 by the fluctuations-- SR is at the origin of the transition. In  large enough systems --such
 that finite-size fluctuations can be neglected-- and for intermediate field strengths, they
 experience  a dynamical phase transition through a nucleation process 
\cite{Korniss2002Absence}. For stronger forcing, the transition is forced by the exogenous field, with
 no contribution from endogenous factors.

In this paper, we investigate the behavior of systems composed of many interacting constituents under
 the influence of a time-varying external forcing. The Ising model framework is used as a generic
 example of such systems. In contrast to the classical SR studies, where the period of the periodic
 forcing is of the order of the Kramers time, we are interested in much faster signals. Additionally,
 aperiodic signals are included to the external forcing, a setup much closer to real world examples,
 which will yield some surprising differences to the cases involving periodic forcing.

We find, for periodic and aperiodic signals alike, that for intermediate values of the
 noise intensity, the system dynamics shows a maximum in amplitude
 \cite{Jung1992Collective,Leung1999Nontrivial} .
 Interestingly, the phenomenon of increased amplitude, which consists of an amplification of
 the signal for a periodic forcing, morphs into an increase of the system-wide fluctuations, 
 uncorrelated with the signal for an aperiodic forcing. We call this phenomenon
 {\em ``noise-induced volatility''} (NIV).

There are many examples of systems composed of many interacting units that are subjected to a rapidly varying --periodic or aperiodic-- common forcing.
 A first example refers to the empirical observations of strong amplifications of thermal noise into effective
 renormalized temperatures by quenched heterogeneities in materials  \cite{Ciliberto2001Effect},  
 in organized flows in liquids \cite{CrossHohenberg93}  
 and in granular media near jamming \cite{Onoetal02}. We argue that NIV
 also provides a conceptual framework to model the immune systems of complex biological organisms,
 viewed as multi-stable complexes, which switch their mode
 of operation under the influence of noisy perturbations by pathogens and other stress factors
 \cite{PandeyStauffer90,BewickZhangSharon09,SoretalendoIS09}.

 Another important application of the proposed mechanism of volatility amplification can be
 found in financial markets. The phenomenon of ``excess volatility"  \cite{Shiller1981Do}
 constitutes one of the major unsolved puzzles in financial economics and refers to the
 ubiquitous observations that financial prices fluctuate with much larger amplitudes than
 they should if they obeyed the fundamental valuation formula, linking the share price of a company
 to it's expected future dividends and discount factors \cite{Brealeyetal}.
 The model described below can be applied to represent a market of interacting investors, where
 the external forcing represents the news, i.e. the publicly available information about
 the traded assets, that investors use to update their estimates of the asset's fair value.
 In addition to the phenomenon of the increased volatility compared to the news amplitude, our framework allows us to address two
 other well known phenomena of financial markets. Namely the fact that
 the news are poor predictors of future price changes \cite{cutler-1989} and the phenomenon of clustered volatility,
 quantified by the slowly decaying temporal dependence of volatility \cite{Ding1993Long}.

We document the phenomenon of noise-induced volatility by numerical and theoretical calculations
 on a stochastic dynamical version of the Ising model on fully connected, regular as well as random
 networks, in the presence of rapidly varying periodic and aperiodic signal.
 NIV also constitutes a new indicator for an approaching phase transition \cite{Scheffer2009Earlywarning}.

This paper is organized as follows. In the following section, we introduce the model studied and
 the measures chosen to quantify the phenomenon studied.
 In Section \ref{sec:periodic}, we revisit the case where the system is driven by a periodic forcing,
 focusing on the case of fast signals, by means of Monte-Carlo simulations and by means of an
 analytical approach. In Section \ref{sec:aperiodic}, we present the main contribution of the paper:
 the study of the system driven by an aperiodic forcing. In Section \ref{sec:excess}, we go
 beyond the fully connected case, focusing on different network topologies and on a paradigmatic
 example of this phenomenon: the excess volatility in financial markets. Finally, Section
 \ref{sec:conclusions} presents a discussion and conclusions of the obtained results.

\section{Model description and diagnostic variables}

Consider a system composed of $N$ interacting units that can be in one of two states: $s=\pm 1$.
The units are updated sequentially, randomly chosen at each unit micro-time $\delta=1/N$,
 i.e.~$N$ updates are equivalent to one time unit at the macroscopic level. The update
 of the state $s_i$ of a given unit $i$ from $t$ to $t+\delta$ is given by
\begin{equation}
s_i(t+\delta) = \textrm{sign}\left(
f(t) + \xi_i(t) + K(t)\sum^N_{j=1}\omega_{ij}\, s_j(t) \right).  \label{eq:opinion-update}
\end{equation}
The value $s_i(t+\delta)$ is determined by three competing contributions:
 (i) a common external dynamic forcing term $f(t)$ (force, pathogens abundance, news);
 (ii) an annealed
 unit-specific term $\xi_i(t)$ that we will call {\em noise} (thermal fluctuations or threshold, intrinsic
 susceptibility of a unit immune system compartment, investor idiosyncratic opinion or private information);
 (iii) an interaction term between units controlled by the amplitude $K(t)$
 (elastic coupling, feedback loops between immune system elements, social impact).

The system's behavior will be investigated under the influence of two different types
 of external signals. To relate to the existing literature, we will use a smooth periodic signal,
 $f_p(t) =A \sin(\omega t)$, with period $2 \pi/\omega$ and strength $A$; this implies that, when averaged over time, the
 standard deviation of the signal is $\sigma_{f_p}=A/\sqrt{2}$. Along this paper, we denote by 
 {\em signal amplitude} the standard deviation of the signal, $\sigma_{f}$.
 As a periodic signal is a rather stylized and artificial setup,
 we will, in a later Section, also analyze the response of such a system to a stochastic process.
 The simplest choice of a stochastic process with tunable characteristic timescale is the
 Ornstein-Uhlenbeck (OU) process, which has exponentially decaying memory and is defined by
 $df_{ap} = -\theta f_{ap} dt + A\, dW_t$, with 0 mean, strength $A$, inverse time-scale $\theta>0$
 and $W_t$ is a Wiener process with normalized variance and zero mean. The asymptotic
 solution of the OU-process is
\begin{equation}
f_{ap}(t) = A\int_{-\infty}^t e^{-\theta(t-\tau)}\,dW_{\tau}, \label{eq:OU_signal}
\end{equation}
which gives a signal amplitude $\sigma_{f_{ap}}=A/\sqrt{2\theta}$.

The noise term $\xi_i(t)$ of each unit in Eq.~\eqref{eq:opinion-update}
 follows an independent stochastic process, whose values are drawn from the cumulative
 distribution function $G(0,D)$, with zero mean ($\average{\xi_i(t)} = 0$) and variance $D^2$.
 Thus, $\average{\xi_i(t)\,\xi_j(t')} = D^2\delta(t-t') \delta_{ij}$.
 If $f(t)=0$ and $G(0,d)$ corresponds to a Logistic distribution, the
 dynamical rule of Eq.~\eqref{eq:opinion-update} is
 equivalent to the kinetic Ising model with Glauber dynamics (cf. appendix) where $D^2$
 is related to the temperature.

In the interaction term in Eq.~\eqref{eq:opinion-update}, the
matrix of weights $\omega_{ij}$ defines the network connectivity between units,
both in topology and relative strength. We assume that the interactions between units are
governed by connections that evolve much slower than the dynamics of the whole system.
This amounts to considering a static network with fixed normalized weights
$\sum_j \omega_{ij} = 1$. The effective coupling strength 
is given by $K(t)$, which may depend on time to reflect global softening-hardening in
rupture processes,  evolving physiological states of immune systems and changes of
social cohesiveness and/or social influence in financial markets.

The macroscopic dynamics of the system is captured by the instantaneous ``magnetization''
\begin{equation}
m(t)= \frac{1}{N}\sum_i s_i(t), \label{eq:magnetization}
\end{equation}
which fluctuates around its mean value 
\begin{equation}
Q=\avgt{m(t)}=\frac{1}{T}\int_0^Tm(t)dt,  \label{eq:Q_simulationAVmag}
\end{equation}
where $T$ is the duration of the simulation. We study the normalized standard deviation
\begin{equation}
\tilde{\sigma}=\frac{\sigma_r}{\sigma_f} = \frac{\sqrt{\avgt{ \left( m(t) - Q\right) ^2}}}{\sqrt{\avgt{ f(t) ^2}}}, \label{eq:normalized_std}
\end{equation}
of $m(t)$, describing the ``volatility'' of the system dynamics scaled by the signal amplitude, $\sigma_f$.
 The correlation between the input signal and the magnetization, defined by
\begin{equation}
\rho(\tau) = \frac{\avgt{ \left( r(t+\tau) - Q \right) f(t) }}{  \sigma_r \sigma_f} ,\label{eq:correlation}
\end{equation}
 provides an additional insight on the level of synchronization between the external influence
 and the overall system dynamics at a lag of $\tau$. The lag where the correlation is
 maximal will be called optimal lag, $\tau^* = \max_\tau\rho(\tau)$.

In the case of the periodic signal, a common measure in stochastic resonance research is the
 spectral amplification factor (SAF)\cite{Jung1991Amplification},
\begin{equation}
 R = \frac{S_\omega[m(t)]}{S_\omega[f_p(t)]} = \frac{S_\omega[m(t)]}{\sigma_f^2},\label{eq:SAF}
\end{equation}
which is the ratio of the power spectrum density of the magnetization $S_\omega[m(t)]$
 over the power spectrum density of the driving signal, $S_\omega[f_p(t)] = \sigma_f^2 = A^2/4$, both
 at the driving frequency $\omega$.

\section{Periodic signal}\label{sec:periodic}
\subsection{Simulations results}

\begin{figure}
\centering
\includegraphics[width=0.45\textwidth,clip]{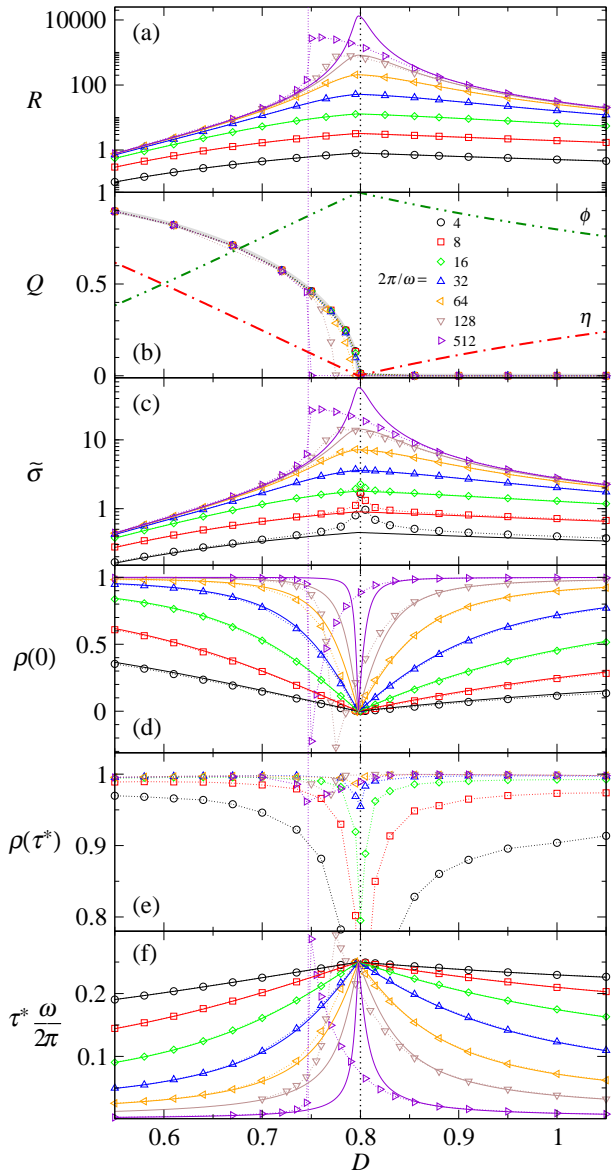}
\caption{For periodic signals: different measures as function of the standard deviation of the noise in
 Eq.~\eqref{eq:opinion-update}, $D$, for weak periodic signals with amplitude
 $\sigma_{f_p}=0.01$. The period of the signal is in the legend of (b).
 Symbols are obtained for simulations with a system size $N=10^6$ and the lines are the
 result of the linear approximation presented in the main text. The
 vertical thin dotted line indicates $D_T$ for $\omega=2\pi/512$, where the transition
 from intra- to inter-well dynamics occurs. The vertical thick 
 dotted line indicates the critical value of noise, $D_c$, where the phase transition
 takes place. For $D\in]D_T,D_c[$, $m(t)$ performs inter-well dynamics.
 \textbf{(a)} Spectral amplification factor, $R$, as defined by Eq.~\eqref{eq:SAF}.
 \textbf{(b)} Average magnetization, $Q$. For $D<D_c$, $Q=0$ indicates that the system performs 
 inter-well dynamics.
 \textbf{(c)} The normalized standard deviation, $\tilde{\sigma}_p$.
\textbf{(d)} Instantaneous correlation between $m$ and $f$, $\rho(0)$, defined by Eq.~\eqref{eq:correlation}.
\textbf{(e)} Correlation between $m$ and $f$ at the optimal lag.
\textbf{(f)} Optimal lag, $\tau^*$, normalized by the period of the driving force.}
\label{fig:periodic:MF-measures}
\end{figure}

First, we consider an homogeneous, complete, network ($\omega_{ij}=1/(N-1)$) and a constant coupling
 strength $K(t)=k=1$. The results reported below are not significantly different for random
 graphs with large average connectivity or when the connections allow for an unbiased
 statistical sampling within the population.
 As previously said, we set $G$ to be a Gaussian distribution with standard deviation $D$, and zero mean.
 Even though the system loses it's equivalence 
 to the kinetic Ising model with Glauber dynamics, all the qualitative
 properties of the system remain unchanged. Without external forcing ($A=0$), the
 system experiences, as for the equilibrium Ising model, a continuous phase transition at $D_c\simeq0.80k$,
 separating the ordered phase with two stable fixed points at $\pm Q(D)$, from the disordered
 phase, with a single stable fixed point at $Q(D)=0$. For the equilibrium case ($A=0$),
 the dependence of $Q(D)$ as function
 of $D$ is shown by the continuous line in Fig.~\ref{fig:periodic:MF-measures}(b).

In Fig.~\ref{fig:periodic:MF-measures}(a),
 we plot the spectral amplification factor as a function of noise strength for signals with
 different periods. The symbols are obtained by simulations of the model with
 $10^6$ units.
 We observe that, even for relatively small periods, an increase of amplification exceeding one
 order of magnitude is achieved for a broad range of intermediate value of noise, the
 hallmark of stochastic resonance.
 Fig.~\ref{fig:periodic:MF-measures}(b) shows the average magnetization, $Q(D)$,
 which is the usual order parameter in the kinetic Ising model studies, for the same signals as in
 Fig.~\ref{fig:periodic:MF-measures}(a).
 In section \ref{sec:MF}, we will develop a theory which shows that the global dynamics can be
 assimilated to a motion within a potential that exhibits a transition from a monostable to a bistable
 regime.
 For fast signals, $Q(D)$
 has a value close to the equilibrium fixed point (continuous line), suggesting that the dynamics
 of $m(t)$ can be well described as fluctuations around this stable (or metastable) point,
 i.e.~$m(t)$ performs intra-well dynamics.
 For slower signals, larger fluctuations around the equilibrium fixed point are observed,
 as $Q(D)$ vanishes already for $D<D_c$,
 indicating symmetric oscillations around $m(t)=0$, i.e. $m(t)$ performs inter-well dynamics.
 By $D_T(A,\omega,N)$, we denote the threshold value of the noise strength at which the potential
 barrier between the two equilibrium fixed points become small enough, such that the system performs
 inter-well hopping and thus $Q(D)$ goes to zero. The noise strength threshold $D_T$ approaches 
 $D_c$ as either $\omega$ or $N$ are increased or $A$ is decreased. In the opposite limits, it will tend to zero.

Independent on the driving frequency, the maximum in the amplification is always found at $D_T$.
 For fast signals where $D_T\sim D_c$, this maximum is observed at the equilibrium phase transition.
 This happens in the presence of two competing phenomena near the equilibrium phase transition:
 on the one hand, a divergence in the susceptibility --making the system very sensitive to
 small changes in the external influences--; on the other, critical slowing down,
 which inhibits the reaction of the system.
 For slow signals, where $D_T<D_c$, together with a more abrupt vanishing of $Q$,
 a pronounced jump in the spectral amplification factor $R$, defined in Eq.~\eqref{eq:SAF},
 is observed at $D=D_T$, where the response of the system is
 greatly increased by the transition from intra- to inter-well dynamics. For $D_T<D<D_c$, the
 amplification decreases with $D$, as the position of the minimum ($\pm m_0(D)$) approaches 0
 for $D$ approaching $D_c$ from the left.

Fig.~\ref{fig:periodic:MF-measures}(d) shows the dependence of the correlation
 at zero lag on the noise strength. A minimum of instantaneous correlation is observed
 at the same values of $D$, where the maximum in $R$ occurs. This result confirms the existence
 of a double peak of the non-normalized instantaneous covariance,
 as was found by Leung et al.\cite{Leung1999Nontrivial,Leung1998Response}.

\begin{figure}
\centering
\includegraphics[width=0.45\textwidth,clip]{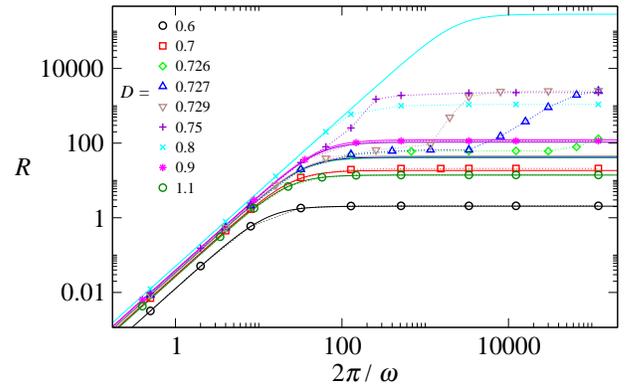}
\caption{For periodic signals with amplitude $\sigma_{f_p}=0.02/\sqrt{2}$ and different values for
 the standard deviation of the noise, $D$, specified in the legend.
 The spectral amplification factor, $R$, as a function of the period, $2\pi/\omega$.
 Symbols are obtained for simulations of a system with $N=9\times 10^4$ and
 the lines represent the linear approximation given by Eq.~\eqref{eq:periodic:MF-SAF}.
 Three regimes of $R$ can be identified as a function of the period: 1) increasing amplification
 with increasing period, 2) plateau with intra-well dynamics, 3) for $D\lesssim D_c$, stark
 increase of amplification due to inter-well dynamics. Similar results are found for larger system
 sizes, where the third regime appears for larger periods as the finite size fluctuations are
 reduced.}
\label{fig:periodic:MF_SAF_VS_period}
\end{figure}

The effect of the signal frequency on the system behavior is shown in
 Fig.~\ref{fig:periodic:MF_SAF_VS_period}, where we plot the amplification $R$ as a function of the period
 of the signal for different values of $D$. For $D>D_c$, where the macroscopic system dynamics is described by
 a monostable potential, the dependence is composed of two regimes. The first regime
 is where the amplitude of the oscillations of $m(t)$
 increases with the period as $m(t)$ is pushed for longer durations into one direction, allowing for
 greater deviations from the origin. For larger signal periods, $R$ reaches a plateau, which constitutes the
 second regime, where the diffusive motion of $m(t)$ is confined by the potential.
 The same behavior is observed for $D\ll D_c$, where the potential barrier between the two 
 minima cannot be overcome by the system, restricting the dynamics to intra-well motions.
 Finally, for $D\lesssim D_c$, the dependence shows a transition into a third regime.
 If the potential barrier is not too high compared with the noise intensity and the finite-size
 fluctuations, the system is able to perform inter-well dynamics for large enough periods
 of the external driving.
 These inter-well dynamics are observed as a second rapid increase in $R$. The period at which this transition happens is the double
 of the Kramers time.

\subsection{Analytical approach}\label{sec:MF}

In order to understand these results, we now develop a mean-field
 theory, which becomes exact in the thermodynamic limit and for weak signal amplitudes.
 As our system is composed of many interconnected  units,
 we can rewrite Eq.~\eqref{eq:opinion-update} by replacing
 the interaction term by the global instantaneous magnetization
 and by explicitly writing down Eq.~\eqref{eq:opinion-update} in the form of
\begin{displaymath}
s_i(t+\delta) = \left\{
\begin{array}{lcr}
+1 &\hbox{if } &  \xi_i(t)  \geq - k\,m(t) - f(t)\\
-1& \hbox{if } &  \xi_i(t)     < - k\,m(t) - f(t) \\
\end{array}
\right. .
\end{displaymath}
Averaging over multiple noise realizations, the expected value for the state of the $i$--th unit,
 at time $t+\delta $ is thus given by
\begin{eqnarray}
\avg{s_i(t+\delta)}_\xi &=& 1-G\big(-k\,m(t)-f(t)\big) \nonumber \\
 & & - G \big(-k\,m(t))-f(t) \big) \nonumber \\
&=&1-2G(-k\,m(t)-f(t)), \label{eq:MF:expected_value_of_magn}
\end{eqnarray}
where $G(\theta)$ is the the cumulative distribution function of the noise term
$\xi_i(t)$, i.e.~$G(\theta)=\int^{\theta}_{-\infty}d\theta'g(\theta')$. Summing over all the units,
 we get that the updated instantaneous magnetization is exactly
\begin{equation}
m(t+\delta)=m(t)+\frac{1}{N}\left[s_i(t+\delta)-s_i(t)\right]. \label{eq:MF:magnetization_update_1}
\end{equation}
By identifying  $1/N = \delta  $ as $dt$, averaging over the complete population
 and taking the continuous limit, the dynamics reads
\begin{equation}
\left\langle\frac{dm(t)}{dt}\right\rangle=\left\langle s_i(t+dt)\right\rangle - m(t). \label{eq:MF:magnetization_update_2}
\end{equation}
With $\dot m(t) = \left\langle{dm(t)}/{dt}\right\rangle$ and substituting
 Eq.~\eqref{eq:MF:expected_value_of_magn} into Eq.~\eqref{eq:MF:magnetization_update_2}, we get
\begin{eqnarray}
\dot{m}(t)&=& -m(t) +1 -2G\big(-k\,m(t)-f(t)\big),\label{eq:MF:r^dot1}
\end{eqnarray}
which constitutes a general closed form evolution equation for the magnetization of the system.

From the point of view of the mean-field limit, the noise $\xi_i(t)$ can
 be either quenched or annealed, as the complete noise is condensed into the last term in
 Eq.~\eqref{eq:MF:r^dot1}.
 For $A=0$ (no external driving), the stationary solution of Eq.~\eqref{eq:MF:r^dot1} gives the dependence
 of the equilibrium fixed point $m_0(D)$ as the solution of the implicit equation
\begin{equation}
 m_0(D) = 1 -2G(-k \,m_0(D)). \label{eq:m_0}
\end{equation}
This solution exhibits a supercritical pitchfork bifurcation as a function of $D$, as
 expected for an Ising-like system, which is displayed by the continuous line
 in Fig.~\ref{fig:periodic:MF-measures}(b).
 The critical parameter is found equal to  $D_c=k\sqrt{2/{\pi}}$, when $\xi_i(t)$ is
 drawn from a Gaussian distribution.

The emphasis of this paper is on the system's reaction to fast, sub-threshold ($A \ll 1$) signals, so
 that  inter-well dynamics can be neglected. Thus, a perturbation expansion $m(t) = m_0+ m_1(t)$ up to first order yields
\begin{equation}
\frac{d}{dt} m_1(t) = - \eta(D)\, m_1(t) + \phi(D)\, f(t) + \mathcal{O}(m_1^2),  \label{eq:MF:r^dot}
\end{equation}
where $\phi(D) \equiv 2 g(- k m_0)$, $\eta(D) \equiv  1- 2 k g(-k m_0)$ and $g=dG/d\xi$.
 The dependence of $\phi$ and $\eta$ as a function of the noise strength is
 displayed by the dash-dotted lines in Fig.~\ref{fig:periodic:MF-measures}(b).
 Based on Eq.~\eqref{eq:MF:r^dot}, $\phi(D)$ can be interpreted as the attenuation of
 the signal by the noise in the individual constituents of the system as $\phi(D)\leq1/k$ for any $D$.
 The parameter $\eta(D)$ can be understood as the strength of the restoring force
 that tends to bring $m(t)$ back to its equilibrium value, $m_0$, after being driven away by
 the influence of $f(t)$. The larger $\eta$, the closer the dynamics of $m(t)$ will be
 to $m_0$ and the shorter will be the memory of $m(t)$.
In the particular case where $f(t)$ is constant, $m_1(t)$ approaches the fixed point $f\, \phi(D)/\eta(D)$.
 Since $\phi(D)$ remains finite when $D$ passes through $D_c$, it is the vanishing of $\eta(D)$ at $D=D_c$
 and its smallness in the vicinity of $D_c$ that is at the origin of the amplified volatility.
 Based on Eq.~\eqref{eq:MF:r^dot}, we can now compute the approximate value of the different
 measures for the external signals and compare them to the simulations of the actual
 system.

For the periodic forcing, the dynamics of the magnetization yields
\begin{equation}
 m_p(t)=\frac{A \, \phi}{\eta^2+\omega^2}\left[ -\omega \cos(\omega t) + \eta \sin(\omega t) \right]+m_0. \label{eq:periodic:r}
\end{equation}
Together with Eq.~\eqref{eq:SAF}, this gives a spectral amplification factor equal to
\begin{equation}
R_p = \frac{4}{A^2}\left(\frac{A \, \phi}{\eta^2+\omega^2}\right)^2 \frac{\omega^2+\eta^2}{4} = \frac{\phi^2}{\eta^2+\omega^2}.   \label{eq:periodic:MF-SAF}
\end{equation}
Fig.~\ref{fig:periodic:MF-measures}(a)  shows that, for fast signals (where $D_T\simeq D_c$),
 the value of $R$ obtained from this approximation matches well with the simulation results.
 Deviations from the  approximation appear for slower signals when $D_T$ does not coincide with $D_c$
 and non-linear effects cannot be neglected anymore.

As it can be seen in
 Fig.~\ref{fig:periodic:MF-measures}(a), the spectral amplification factor can
 reach values above 100, showing that this system,  even without considering inter-well dynamics,
 is able to show remarkable reactions to a weak forcing.
Two distinct amplification mechanisms of sub-threshold periodic signals can be identified 
 by comparing the simulation with approximation results. The first mechanism, being present for
 finite and infinite systems, is the increase of the output amplitude by the decrease of the value of $\eta$:
 by reducing the restoring force of $m(t)$, such that it can be further displaced from $m_0$, the
 oscillation amplitudes are increased.
 The second mechanism is the amplification through inter-well jumps, which is only present in finite
 systems, as a sub-threshold driving force cannot overcome the potential barrier without the existence of a source of fluctuations, like finite-size effects.
 Note that in the thermodynamical limit, where the approximation is valid for any frequency,
 at $D_c$  where $\eta(D_c)=0$, it follows from Eq.~\eqref{eq:periodic:MF-SAF} that the fluctuations of $m(t)$
 for nonzero frequencies will always be finite.

In addition to the spectral amplification factor, $R$, we also measure the normalized
 variance of $m(t)$, $\tilde{\sigma}$, which
 measures the volatility of the dynamics, independent of the exact shape of the power spectrum. This
 measure is convenient as it can be used for comparison with the aperiodic signal,
 for which $R$ is not defined. From Eq.~\eqref{eq:periodic:r}, the normalized variance of $r(t)$ is
\begin{eqnarray}
 \tilde{\sigma_p}^2 = \frac{2}{A^2}\,\avgt{m_p(t)^2} &=& \frac{2}{A^2}\,\frac{\omega}{2 \pi} \int_0^{\frac{2 \pi}{\omega}} m_p(t)^2 dt \nonumber\\
&=& \frac{\phi^2 }{\eta^2 + \omega^2}. \label{eq:periodic:normalized_variance}
\end{eqnarray}
The equivalence between the eqs.~\eqref{eq:periodic:MF-SAF} and \eqref{eq:periodic:normalized_variance}
 is due to the use of a linear
 response approximation in the macroscopic dynamics of the systems, neglecting the response at higher order harmonics of the driving signal.
 As a consequence, the approximation of the spectral amplification factor, $R_p$,
 is better fitted by the simulations than $\tilde{\sigma}_p$.
 
From Fig.~\ref{fig:periodic:MF-measures}(c), we see that the mean-field approximation matches well
 the values of $\tilde{\sigma}$ obtained by simulations for intermediate signal periods. For large periods,
 the inter-well dynamics destroys the match, and for fast signals (small periods), the
 finite-size fluctuations overshadow the fluctuations induced by the signals.

The correlation between $m_p(t)$ and $f_{p}(t)$ is given by
\begin{equation}
\rho_p(\tau) = \frac{\eta \cos(\omega \tau)+\omega \sin(\omega\tau)}{\sqrt{\eta^2+\omega^2}}  \label{eq:periodic:cross-correlation}
\end{equation}
and, for the optimal lag, we obtain
\begin{equation}
\tau^*_p = \frac{\arctan(\frac{\omega}{\eta})}{\omega},  \label{eq:periodic:optimal_lag}
\end{equation}
which follows directly from Eq.~\eqref{eq:periodic:r}. The correlation at zero lag is
 shown is Fig.~\ref{fig:periodic:MF-measures}(d).
 As for the results of $R$, the simulation results are well captured by the mean-field approximation for
 high frequencies and deviate due to inter-well dynamics for lower frequencies. As was observed
 in \cite{Leung1999Nontrivial}, a dip in correlation is observed for intermediate
 values of the noise amplitude. This dip occurs at $D_T$, concomitant with the maximum in the
 amplification measured by $R$.
This apparent contradiction can be explained by the results
 shown in Fig.~\ref{fig:periodic:MF-measures}(f), which plots the optimal
 lag between $m(t)$ and $f(t)$ normalized by the period of the signal. At $D_T$, $m(t)$ and $f(t)$ are maximally
 lagged, reducing the correlation at lag zero. On the other hand, the correlation at the optimal
 lag has a value close to one, explaining the high amplification of the signal.
 This behavior can also be found in the approximation, although there the maximum of amplification
 and minimum of instantaneous correlation is found at $D_c$ as our approximation neglects
 inter-well dynamics. From Eq.~\eqref{eq:periodic:cross-correlation}
 and Eq.~\eqref{eq:periodic:optimal_lag} it follows that, in the thermodynamic limit,
 there exists an optimal lag, $\tau^*_p$, for which perfect correlation is achieved for
 any frequency and any noise strength, i.e. $\rho_p(\tau^*_p)=1$.

\begin{figure}
\centering
\includegraphics[width=0.45\textwidth,clip]{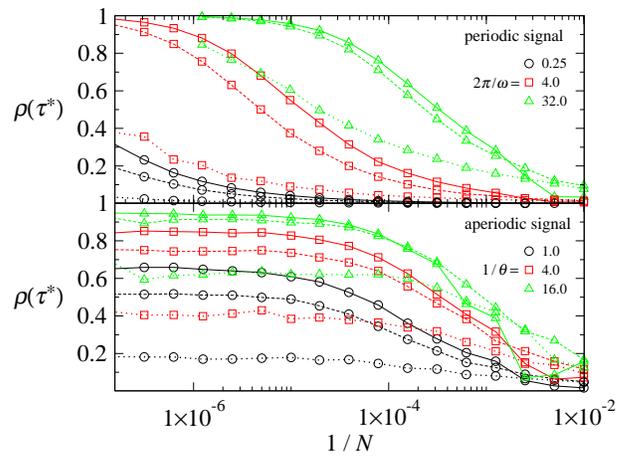}
\caption{Size dependence of the maximal correlation between the magnetization and the signal
 with $\sigma_f=0.01$ and different time-scales for several noise
 intensities: $D=0.7$ (full lines), $D=0.8$ (dotted lines), $D=0.9$ (dashed lines).
 Characteristic time-scales of the signal are specified in the legends.
\textbf{(top)} Periodic signals, where $\rho(\tau^*)$ converges to 1 for any value of $D$ in the
 thermodynamic limit. The rate of convergence increases with the signal period.
\textbf{(bottom)} In case of aperiodic signals, the value of $\rho(\tau^*)$ does not converge 
 to 1 in the thermodynamic limit and depends on $D$ and $\theta$. At $D=D_c$, the maximal
 correlation, $\rho(\tau^*)$, is significantly reduces compared to $D\neq D_c$.}
\label{fig:periodic+OU:MF_MaxCorr_VS_divStd_and_size}
\end{figure}

Perfect correlation is however not achieved for any frequency in finite systems as shown by the dependence of
 $\rho(\tau^*)$ in Fig.~\ref{fig:periodic:MF-measures}(e),
 where the deterioration of the correlation with increasing driving frequencies is
 observed. The origin of this effect lies in the finite-size fluctuations, which vanish in the
 thermodynamic limit, as Fig.~\ref{fig:periodic+OU:MF_MaxCorr_VS_divStd_and_size}(top) shows.
 $\rho(\tau^*)$ converges to 1 for infinite systems, at a rate of convergence depending
 on $D$ and the driving period.


\section{Aperiodic Signal}\label{sec:aperiodic}

\begin{figure}[ht]
\centering
\includegraphics[width=0.45\textwidth,clip]{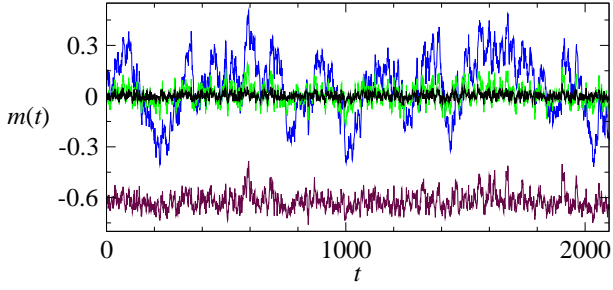}
\caption{For aperiodic signals: Time evolution of the magnetization, $m(t)$, for different noise intensities $D$
 obtained with the same realization of the driving force $f(t)$ and $\xi_i(t)$.
 $N=10^4$,  $\sigma_{ap}=0.04$, $\theta=1.0$,  $k=1.0$, $D = 2.0, 1.0, 0.8$ (smaller to larger amplitude of
 $m(t)$'s fluctuations) and $D=0.7$ (bottom curve fluctuating around $m(t) =-0.6$).}
\label{fig:OU:MF-realizations}
\end{figure}

\begin{figure}[ht]
\centering
\includegraphics[width=0.45\textwidth,clip]{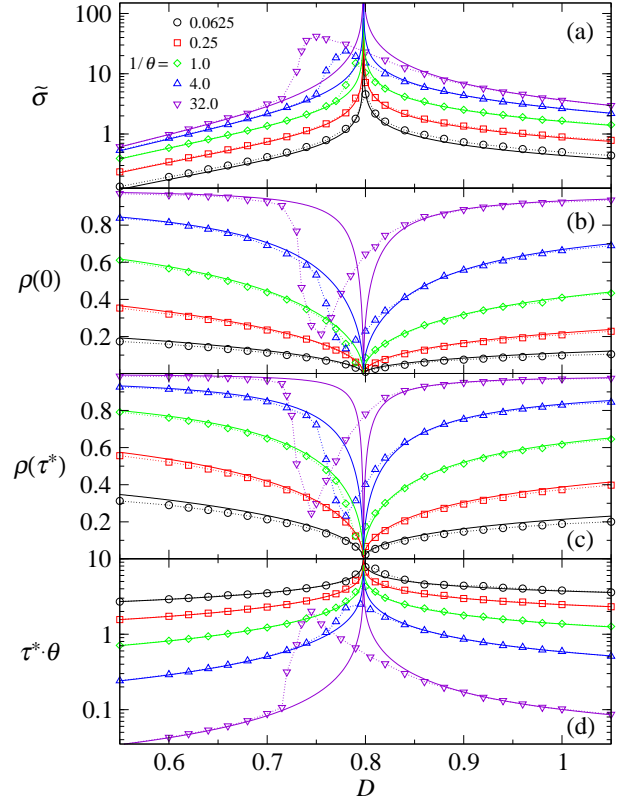}
\caption{For aperiodic signals: different measures as function of the standard deviation of the noise in
 Eq.~\eqref{eq:opinion-update}, $D$, for weak aperiodic signals with amplitude
 $\sigma_{f_p}=0.01$. The inverse of the signal's time-scale, $\theta$, is given in the legend.
 Symbols are obtained for simulations with a system size $N=10^6$ and the lines are the
 result of the linear approximation presented in the main text.
 \textbf{(a)} Normalized standard deviation, $\tilde{\sigma}$, measuring the volatility amplification.
 \textbf{(b)} Instantaneous correlation between $m$ and $f$, $\rho(0)$, defined by Eq.~\eqref{eq:correlation}.
 \textbf{(c)} Correlation between $m$ and $f$ at the optimal lag.
 In contrast to the periodic signal, a system driven by an aperiodic signal
 is not able to follow the signal, even at the optimal lag, which is well described by the mean-field approximation.
 \textbf{(d)} Optimal lag, $\tau^*$, normalized by the time-scale of the signal.}
\label{fig:OU:MF-measures}
\end{figure}

We now turn our attention to the case where the common forcing is aperiodic. In this section,
 we will consider an external force, which is described by the OU-process introduced in Eq.~\eqref{eq:OU_signal}.
Fig.~\ref{fig:OU:MF-realizations} illustrates the typical dynamic
 behaviors of $m(t)$ for different values of noise strengths $D$ for a single realization of the
 driving force $f_p(t)$. For $D\geq D_c$, the magnetization fluctuates around $m_o(D)=0$, with
 increasing amplitudes as $D$ approaches $D_c$. The fluctuation amplitude deceases again when $m(t)$
 performs intra-well dynamics for $D<D_c$, where $Q(D)\neq0$.

 We will compute
 the same observables as for the periodic signal and investigate the differences. The formal
 solution of the linearized version of the dynamics, eq.~\eqref{eq:MF:r^dot}, is given by
\begin{equation}
 m_1(t) = A \, \phi \int_{-\infty}^t e^{-\eta(t-\tau)}\int_{-\infty}^\tau e^{-\theta(\tau-\tau')}\,dW_{\tau'}\,d\tau, \label{eq:OU:r}
\end{equation}
which describes the dynamics around $m_0$.
 The normalized variance of $m_1(t)$ described by eq.\eqref{eq:OU:r} is now given by
\begin{equation}
\tilde{\sigma}_{ap}^2 = \frac{\phi^2}{ \eta\,(\theta + \eta)}. \label{eq:OU_normalized_variance}
\end{equation}
In Fig.~\ref{fig:OU:MF-measures}(a), we compare the normalized variance obtained by
 means of numerical simulations with this theoretical result. We find a very good agreement between
 the two for fast signals, with the same deficits due to inter-well dynamics as for the periodic
 case with slower signals.
 However, by comparing $\tilde{\sigma}_{ap}$ for the aperiodic signal with the
 periodic case, $\tilde{\sigma}_{p}$, we observe a major difference. Whereas for the periodic case,
 the volatility of the dynamics shows a finite maximum value at $D_c$,
 the volatility diverges if the system is driven by an aperiodic signal as $\eta(D_c)=0$.
 This divergence of the normalized volatility, $\tilde{\sigma}_{ap}$, is not to
 be understood as an explosion of the dynamics, as $m(t)$ cannot exceed $[-1,+1]$. It reflects
 the immensely amplified reaction to a weak external forcing, consistent with the diverging 
 susceptibility in equilibrium phase transitions. The fact that this divergence is absent
 for a periodic forcing stems from the discreteness of the power spectrum of the input signal.

The correlation between the forcing $f_{ap}(t)$ and the magnetization, $m(t+\tau)$, is given by
\begin{equation}
\rho_{ap}(\tau) = \frac{\sqrt{\eta^2 + \theta\eta}}{ \eta^2 - \theta^2 } \Big[(\eta+\theta)\, e^{-\theta\tau}-2 \theta \, e^{-\eta\tau}\Big] \label{eq:OU:normalized_correlation}
\end{equation}
and is shown in Fig.~\ref{fig:OU:MF-measures}(b) for zero lag, together
 with the simulation results, which are found in good agreement. The optimal lag for which $\rho_{ap}(\tau)$ is
 maximum occurs at
\begin{equation}
 \tau^*_{ap}=\frac{\ln(\frac{\eta+\theta}{2\eta})}{\theta-\eta} \label{eq:OU_optimal_lag}
\end{equation}
yielding a maximal correlation of
\begin{equation}
 \rho_{ap}(\tau^*_{ap})=2^\frac{\theta}{\theta-\eta} \, \left(\frac{\eta}{\theta+\eta}\right)^{\frac{\eta+\theta}{2(\theta-\eta)}}. \label{eq:OU:maximal_covariance}
\end{equation}
Here, we find the second major difference between the periodic and aperiodic driving. For the
 periodic signal, it is always possible to find a lag at which the correlation between the forcing
 and the system's response is perfect. For the aperiodic signal (see Fig.~\ref{fig:OU:MF-measures}(c)
 and Fig.~\ref{fig:periodic+OU:MF_MaxCorr_VS_divStd_and_size}(bottom)),
 on the other hand, even in the thermodynamic limit, the dynamics of the system can be almost
 unrelated to the forcing.
 Perfect correlation is only reachable for very slow signals, i.e.~
 $\displaystyle\lim_{\theta\to\infty}\rho_{ap}(\tau^*_{ap}) = 1$. Given that the forcing --an OU-process-- has
 a continuous power spectrum, and that the response of the system is frequency-dependent, the spectrum
 of the macroscopic dynamics is distorted when compared to the one of the forcing, which has the effect of 
 decreasing the correlation. Indeed, the system is only able
 to follow the part of the signal spectrum with frequencies lower than $\eta(D)$, which describes
 the rate at which the system can effectively react to external stimuli. As with decreasing
 $\theta$, the contribution of lower frequencies in the signal's spectrum is higher, the correlation
 for fixed $D$ increases with decreasing $\theta$.

For $D \approx D_c$, the volatility amplifies many times that of the driving signal $f(t)$.
 Concomitantly, $\rho$ vanishes for every value of the lag $\tau$, indicating that the volatility
 of the system is generated by an internal collective behavior. It is important to note that,
 even though the system dynamics are endogenously generated, they are initiated by an exogenous
 driving of the system. This is further confirmed by the good agreement between the approximation and
 simulations for fast signals.
 It is the shadow of the diverging susceptibility together with the vanishing rate of the reaction
 of the equilibrium model at $D_c$, which is responsible for the observed NIV phenomenon.

\section{Extensions of the phenomenon studied} \label{sec:excess}
\subsection{Different Networks} \label{sec:different_networks}

\begin{figure}
\centering
\includegraphics[width=0.45\textwidth,clip]{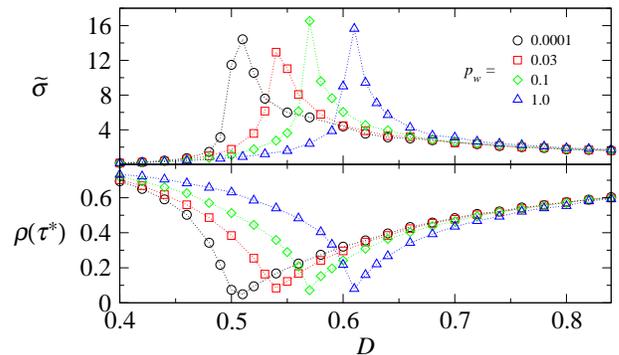}
\caption{For an aperiodic signal: \textbf{(top)} the normalized volatility
 $\tilde{\sigma}$ and \textbf{(bottom)} the correlation at optimal lag, $\rho(\tau^*)$,
 as function of the standard deviation of the noise, $D$, for
 different small-world random connection concentrations $p_w$ of a two-dimensional regular
 grid with Moore neighborhood.
 For $p_w=0$, the network is a 2D regular grid with Moore neighborhood.
 For $p_{w}=1.0$, the network is a random graph with an average degree, $d=4$.
The other system parameters are $N=10^6$, $\sigma_f=0.01$, $\theta=1$, $k=1$.}
\label{fig:SW2D-std+Xcorr}
\end{figure}

To show that the NIV-phenomenon, characterized by the increase of volatility and decrease of correlation
 to the aperiodic forcing, is robust with respect to the structure of the network,
 Fig.~\ref{fig:SW2D-std+Xcorr} shows the normalized volatility $\tilde{\sigma}$ and the maximum
 in correlation $\rho(\tau^*)$ as a function of $D$ (the standard deviation of the noise term in Eq.~\eqref{eq:opinion-update})
 for different networks. We consider a two-dimensional regular  grid with Moore neighborhood
 and random small-world connections with varying concentration $p_w$.
 Changing $p_w$ from $0$ to $1$ interpolates between the regular 2D lattice and the completely
 random network. For each $p_w$, the peak in volatility is still concomitant with the vanishing of $\rho$
 at the threshold value $D_T(p_w)$. The noise intensity threshold $D_T(p_w)$ is increasing in $p_{w}$,
 as larger global interconnection enhances the cooperative organization, and
 larger noise is needed to destroy the ferromagnetic state.

\subsection{Excess volatility in financial markets}

By the definition of our model given by Eq.~\eqref{eq:opinion-update}. it is clear that
 it is also interpretable as a model of opinion dynamics, where $s_i(t)$ is the opinion of agent $i$
 at time $t$, in line with the established literature on discrete choice \cite{Brock2001Discrete}.
 The external forcing $f(t)$ can be seen as the news that are common to all agents,
 the noise $\epsilon_i(t)$ contains the agents' private information and the coupling
 term represents the social interaction between agents. The dynamics of the global opinion
 is then given by $m(t)$.

When applied to the social system of financial markets, the agents are investors and
 $s_i(t)$ corresponds to their opinion on whether the asset is over- or under-priced and hence
 to their willingness to buy (+1) or to sell (-1). The global demand is the given by $m(t)$, 
 which impacts on the price as
\begin{equation}
 \log [p(t+1)]= \log [p(t)] + \frac{m(t+1)}{\lambda}. \label{eq:price_impact}
\end{equation}
 Here $\lambda$ represents the liquidity depth of the market, which is assumed constant
 and $m(t)/\lambda$ is the financial return, $r(t)$, from period $t$ to period $t+1$.
 This equation expresses a linear market impact of the demands, which is a common hypothesis
 in stylized models of financial markets \cite{Beja1980Dynamic, Wyart2007Selfreferential}.
 The results below do not change qualitatively for more general non-linear
 impact functions \cite{Lillo2003Econophysics}.

To apply our model to the financial markets, we use the coupling strength, $k$, instead of $D$ as the control parameter.
 Rather than assuming a fixed coupling strength for investors, we propose that
 the impact of colleagues' opinions on a given investor may be slowly varying with time.
 This effect reflects the fact that,
 in times of greater uncertainty, investors tend to be more influenceable by their
 surrounding \cite{Bikhchandani1992Theory}.
 There are many varying sources of uncertainty that impact financial markets, including
 the economic and geopolitical climate and past stock market performance.
 In the spirit of Ref.~\cite{Stauffer1999Selforganized}, all these factors are embodied
 into the notion that $K(t)$ undergoes a slow random walk
 with i.i.d.~increments $K(t+\delta t)-K(t)\sim N(0,\sigma_k)$, which is confined in the
 interval $[k-\Delta k; k+\Delta  k]$. This later constraint ensures that social imitation
 remains bounded. We could have used an Ornstein-Uhlenbeck process or
 any other such confining dynamics, without changing the crucial results presented below.
 More complex models of sophisticated investors involve the strategic adaptations of the
 traders' propensity to imitate to the reliability of their colleagues in recent
 outcomes \cite{Sornette2006Importance,Zhou2007Selforganizing,harras2011bubble}.

\begin{figure}
\centering
\includegraphics[width=0.45\textwidth,clip]{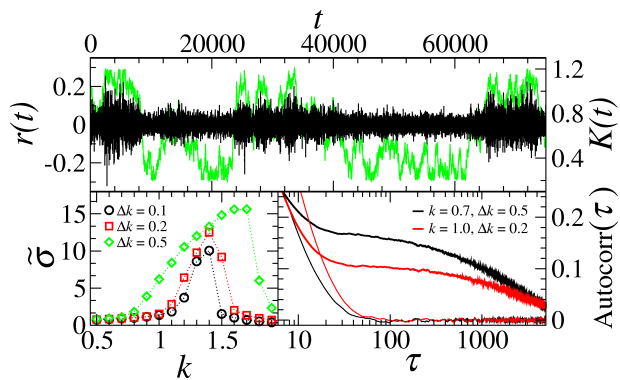}
\caption{(color online) Upper panel: Sample dynamics $r(t)$ (black bursty line) when the coupling
 strength $K(t)$ of the interactions between agents
 undergoes a confined random walk  (green) in $[k-\Delta k; k+\Delta  k]$ with $\Delta k = 0.5$
 and step size $\sigma_k={\Delta k}/{\sqrt{5000}}$.
 Lower-right panel: quickly vanishing (resp. long memory of) the auto-correlation of $r(t)$
 (thin lines) (resp. $|r(t)|$  (thick line)) for two values of $k$ and of $\Delta k$.
 Lower-left panel:  NIV resonance in the presence of a time varying $K(t)$, with
 $\Delta k=0.1$ (circles), 0.2 (squares), 0.5 (diamonds). The other parameters are
 $N=10^4$, $\sigma_{ap}=0.04$, $D=1$.}
\label{fig:MF-dynHERD}
\end{figure}

By the mechanism of sweeping of the coupling strength $K(t)$ close to the critical coupling strength
 $k_c$ (for fixed noise strength $D$)~\cite{sornette1994sweep,Sornette1994Sweeping}, we expect and find transient
 burst of volatility in response to the aperiodic driving force $f(t)$ with constant amplitude and time scale.
 Fig.~\ref{fig:MF-dynHERD} shows a typical
 realization, where the return $r(t)$ exhibits transient bursts,
 associated with excursion of $K(t)$ in the neighborhood of $k_c$. The lower-left panel of
 Fig~\ref{fig:MF-dynHERD} shows the robustness of the NIV phenomenon as a function of the average
 coupling $k$: even with a fluctuating $K(t)$, a large volatility peak appears for intermediate  values of $k$.
 The lower-right panel shows very short-range correlations of $r(t)$ but very long-range
 correlations of the financial volatility $|r(t)|$ (another equivalent proxy
 for volatility), very similar to empirical observations of financial returns \cite{Ding1993Long}.
 Such long persistence of the volatility can be traced back to the slow diffusive nature of
 $K(t)$ in line with the investors' slowly changing trust in the economy.
 From the previous section also follows that during times of crisis and strong social interaction 
 ($k$ close to $k_c$), the dynamics is generated mostly exogenously, well in line with the
 documented inability of news events to explain large price movements \cite{cutler-1989}.

\section{Conclusions} \label{sec:conclusions}

In this paper, we have investigated the behavior of a system composed of  coupled bistable units under the 
 influence of a --rapidly varying-- common exogenous forcing and independent noise sources. Independently of the shape of the driving force, 
intermediate noise strengths trigger a strong level of fluctuations of the macroscopic dynamics 
around the critical value separating the ordered from the disordered phases.
  For a periodic forcing, this peak corresponds to a pronounced amplification of the signal, with a
 strong correlation between the macroscopic dynamics and the driving force at the optimal lag, the paradigmatic signatures of stochastic resonance.

 When the driving force is aperiodic a similar peak appears, but here the amplitude of the fluctuations 
 exceeds by far  those observed for periodic signals. 
 Coincidental with the increase of fluctuations, the correlation between the driving
 force and the system dynamics is completely destroyed. This shows that even though these fluctuations
 are induced by the common forcing, the macroscopic dynamics has an endogenous origin.
 This phenomenon of {\em noise-induced volatility} contrasts with that of stochastic resonance,  
 with the major difference being that it is not the signal, but the fluctuations that are amplified.

Moreover, this phenomenon of {\em noise-induced volatility} also constitutes a new indicator for the approaching of a phase
 transition \cite{Scheffer2009Earlywarning}, and it applies to a broader range of real-world systems
 due to the more common setup given of a coupled system driving by an aperiodic forcing.

As an example of a system where this phenomenon can be observed, we have proposed the social
 system of stock markets, in which we have been able to not only explain the excess of volatility 
 observed  in stock prices, but also the apparent absence of correlation between news and price changes and the 
 persistence of volatility during times of crises.

\appendix*
\section{Equivalence to the kinetic Ising model with Glauber dynamics}

A popular update mechanism in the kinetic Ising model literature was introduced
 by Glauber \cite{Glauber1963TimeDependent}. In it, the probability for a spin to flip is given by
\begin{equation}
p_{\textrm{flip}} = \frac{1}{e^{\beta \Delta E_i}+1} \label{eq:glaubem_1}
\end{equation}
where $\Delta E_i$ is the energy gained by the system through the spin-flip and $\beta=1/kT$.
 With $s_i=\pm1$ and $E_i=-s_i\big(\sum_j K_{ij}s_j+f\big)$, which is the energy of the state $s_i$,
 Eq.~\eqref{eq:glaubem_1} can be rewritten as
\begin{eqnarray}
p_{\textrm{flip}} = p_{s_i\rightarrow -s_i} &=& \frac{1}{e^{s_i 2 \beta \big(\sum_j K_{ij}s_j+f(t)\big)} +1} \nonumber\\
&=& \frac{1}{e^{s_i 2 \beta  \Lambda} +1} \label{eq:glauber_2}
\end{eqnarray}
where $\Lambda = \sum_j K_{ij}s_j+f$. With the transition rate given by
 Eq.~\eqref{eq:glauber_2}, we can compute the probability of being in state $s_i$ at time $t+\delta$ by
\begin{eqnarray}
p(s_i;\,t+\delta) &=& p(s_i;\,t) p_{s_i\rightarrow s_i} + p(-s_i;\,t) p_{-s_i\rightarrow s_i} \\
             &=& p(s_i;\,t) (1-\frac{1}{e^{s_i 2 \beta  \Lambda} +1} ) + \nonumber \\
                & & p(-s_i;\,t) \frac{1}{e^{-s_i2\beta\Lambda}+1} \nonumber \\
             &=& \Big(p(s_i;\,t) + p(-s_i;\,t)\Big) \frac{1}{e^{-s_i2\beta\Lambda}+1} \notag\\
             &=& \frac{1}{e^{-s_i2\beta\Lambda}+1} = p(s_i), \label{eq:glauber_population}
\end{eqnarray}
which is independent of time and gives us the probability of finding spin $i$ in state $s_i$.
 Eq.~\eqref{eq:glauber_population} can be rewritten as
\begin{eqnarray}
s_i(t+\delta) &=& \left\{
\begin{array}{lcl}
+1 &\hbox{with } &  \textrm{Prob} = (e^{-2\beta\Lambda}+1)^{-1} \\
-1 &\hbox{with } &  \textrm{Prob} = (e^{2\beta\Lambda}+1)^{-1} \\
\end{array}
\right. \notag\\
&=& \left\{
\begin{array}{lcl}
+1 &\hbox{with } &  \textrm{Prob} = 1-F(-\Lambda) \\
-1 &\hbox{with } &  \textrm{Prob} = F(-\Lambda) \\
\end{array}
\right.\label{eq:glauber_cases}
\end{eqnarray}
where $F(x)$ is the cumulative density function of a Logistic distribution with zero mean and variance ${\pi^2}/{12\beta^2}$.

The model studied in this paper, defined by Eq.~\eqref{eq:opinion-update}, can be rewritten as
\begin{eqnarray}
s_i(t+\delta) &=& \left\{
\begin{array}{lcr}
+1 &\hbox{if } &   \xi_i(t)  \geq -\Lambda \\
-1 &\hbox{if } &   \xi_i(t)  <    -\Lambda \\
\end{array}
\right. \notag\\
&=& \left\{
\begin{array}{lcl}
+1 &\hbox{with } &  \textrm{Prob} = 1-G(-\Lambda) \\
-1 &\hbox{with } &  \textrm{Prob} = G(-\Lambda) \\
\end{array}
\right.\label{eq:NIV_cases}
\end{eqnarray}
with $G(x)$ being the CDF of the probability density function of $\xi_i(t)$, with zero  mean
 and variance $D^2$. By direct comparison of Eq.~\eqref{eq:glauber_cases}
 and Eq.~\eqref{eq:NIV_cases}, one can see that the model defined by Eq.~\eqref{eq:opinion-update} is
 equivalent to the kinetic Ising model with Glauber dynamics if the distribution of the
 noise is chosen to be a Logistic distribution.



%

\end{document}